\documentclass[aps,pra,twocolumn,superscriptaddress,10pt,nofootinbib]{revtex4-2}
\usepackage{multirow,amsthm,amssymb,amsbsy,amsmath,epsfig,epstopdf,bm,float}
\DeclareMathOperator{\Tr}{Tr}
\usepackage{enumitem}
\usepackage{graphicx}
\usepackage{booktabs}
\usepackage{verbatim}
\usepackage{dcolumn}
\usepackage{color}
\usepackage[dvipsnames]{xcolor}
\usepackage{mathtools}
\usepackage[normalem]{ulem}
\usepackage{soul}
\usepackage[colorlinks=true,linkcolor=blue,citecolor=blue,urlcolor=blue]{hyperref}
\usepackage[export]{adjustbox}
\usepackage{comment}

\makeatletter
\renewcommand{\fnum@figure}{Figure \thefigure}
\makeatother

\newcommand{\ba}{\begin{eqnarray}}
\newcommand{\ea}{\end{eqnarray}}
\newcommand{\be}{\begin{equation}}
\newcommand{\ee}{\end{equation}}
\newcommand{\beq}{\begin{equation}}
\newcommand{\eeq}{\end{equation}}
\newcommand{\bba}{\begin{eqBox}\begin{eqnarray}}
\newcommand{\eba}{\end{eqnarray}\end{eqBox}}

\newcommand{\ket}[1]{\vert #1 \rangle}

\newcommand{\ketbra}[2]{|#1\rangle\langle#2|}

\newcommand{\eg}{{\it{e.g.}~}}
\newcommand{\ie}{{\it{i.e.}~}}

\newcommand{\mean}[1]{\langle#1\rangle}

\newcommand{\overbar}[1]{\mkern 1.5mu\overline{\mkern-1.5mu #1 \mkern-1.5mu}\mkern 1.5mu}

\newcommand{\suppref}[1]{Section {#1} of the Supplementary Information \cite{korhonen2025practical}}

\begin{document}
\title{Practical Techniques for High-Precision Measurements on Near-Term Quantum Hardware and Applications in Molecular Energy Estimation}

\author{Keijo Korhonen}
\email{keijo@algorithmiq.fi}
\affiliation{Algorithmiq Ltd, Kanavakatu 3C 00160 Helsinki, Finland}
\affiliation{QTF Centre of Excellence, Department of Physics, University of Helsinki, P.O. Box 43, FI-00014 Helsinki, Finland.}

\author{Hetta Vappula}
\affiliation{Algorithmiq Ltd, Kanavakatu 3C 00160 Helsinki, Finland}

\author{Adam Glos}
\affiliation{Algorithmiq Ltd, Kanavakatu 3C 00160 Helsinki, Finland}

\author{Marco Cattaneo}
\affiliation{Algorithmiq Ltd, Kanavakatu 3C 00160 Helsinki, Finland}
\affiliation{QTF Centre of Excellence, Department of Physics, University of Helsinki, P.O. Box 43, FI-00014 Helsinki, Finland.}

\author{Zolt\'{a}n Zimbor\'{a}s}%
\affiliation{Algorithmiq Ltd, Kanavakatu 3C 00160 Helsinki, Finland}
\affiliation{QTF Centre of Excellence, Department of Physics, University of Helsinki, P.O. Box 43, FI-00014 Helsinki, Finland.}

\author{Elsi-Mari Borrelli}
\affiliation{Algorithmiq Ltd, Kanavakatu 3C 00160 Helsinki, Finland}

\author{Matteo A.C. Rossi}
\affiliation{Algorithmiq Ltd, Kanavakatu 3C 00160 Helsinki, Finland}

\author{Guillermo Garc\'{i}a-P\'{e}rez}%
\affiliation{Algorithmiq Ltd, Kanavakatu 3C 00160 Helsinki, Finland}

\author{Daniel Cavalcanti}
\affiliation{Algorithmiq Ltd, Kanavakatu 3C 00160 Helsinki, Finland}

\date{\today}

\begin{abstract}
Achieving high-precision measurements on near-term quantum devices is critical for advancing quantum computing applications. Quantum computers suffer from high readout errors, making quantum simulations with high accuracy requirements particularly challenging. This paper implements practical techniques to reach accuracies essential for quantum chemistry by addressing key overheads and noise sources. Specifically, we leverage locally biased random measurements for reducing shot overhead, repeated settings with parallel quantum detector tomography for reducing circuit overhead and mitigating readout errors, and blended scheduling for mitigating time-dependent noise. We demonstrate these techniques via molecular energy estimation of the BODIPY molecule on a Hartree-Fock state on an IBM Eagle r3, obtaining a reduction in measurement errors by an order of magnitude from 1-5\% to 0.16\%. These strategies pave the way for more reliable quantum computations, particularly for applications requiring precise molecular energy calculations.
\end{abstract}

\maketitle

\section{Introduction}
\label{sec:intro}
Quantum computing is poised to revolutionize fields ranging from logistics to life sciences, by solving complex problems that are intractable for classical computers. However, several significant obstacles must be overcome to fully realize the potential of quantum computing. One of the main challenges is the inherent noise present in quantum systems. Noisy gates and readout errors degrade the quality of quantum computations, leading to inaccurate results. Additionally, the low accuracy of measurements, often caused by low statistics due to limited sampling (or “shots”), further compounds the problem. These issues are exacerbated by the inherent resource-intensive nature of the problems that are intended to be run on quantum computers, which are currently limited by the scalability and efficiency of existing quantum computing systems. Thus, enhancing the precision and reliability of quantum measurements is crucial for advancing quantum technology.

In this paper we investigate how a combination of a set of strategies designed to address the primary issues associated with quantum measurements can impact the estimation of energy of a physical system. As a concrete example, we consider a molecular system, namely the Boron-dipyrromethene (BODIPY) molecule, which has been used in applications such as medical imaging, biolabelling, photoelectrochemistry, photocatalysis, artificial photosynthesis, optoelectronics, and photo-dynamic therapy. Simulations of such chemical compounds often aim to evaluate the molecular ground state energy, which can be done by preparing an ansatz state on the quantum computer using \eg the Variational Quantum Eigensolver (VQE) algorithm \cite{peruzzo2014variational, mcclean2016theory}.
The estimation of certain molecular ground state energies on quantum computers has previously been demonstrated experimentally \cite{peruzzo2014variational, kandala2017hardware-efficient,  omalley_scalable_2016, colless_computation_2018, hempel_quantum_2018, robledo-moreno2024chemistry, nützel2025solving}. To obtain the energy of an ansatz state, one needs to evaluate the expectation value of the state and chemical Hamiltonian to a high precision. A commonly used upper bound for this precision, motivated by the sensitivity of reaction rates to changes in energy, is \textit{chemical precision} at $1.6 \times 10^{-3}$ Hartree \cite{cao2019quantum}. For our measurement techniques, we use the term \textit{chemical precision} instead of the more commonly used \textit{chemical accuracy}, to distinguish between the statistical precision in the estimation procedure (\eg the estimation of an ansatz state energy) and the exact error of an ansatz state to a target energy of a molecule (\eg the exact ground state energy of a molecule). The estimation of energies to chemical precision will be also a relevant challenge even in the age of fault-tolerant quantum computing due to the complexity in the observables that represent chemical molecules. In this work, we focus on decreasing the shot overhead (\ie the number of times the quantum computer is measured), the circuit overhead (\ie the number of times one needs to implement a different set of gates on the computer), as well as static and time-dependent measurement noise effects. Each of these issues poses significant obstacles to the accuracy and efficiency of quantum computations, and our strategies offer targeted solutions to mitigate their impact. We show that with the help of all of the aforementioned techniques, we are able to reach estimation errors close to chemical precision on current quantum hardware despite high readout errors on the order of $10^{-2}$ as well as the complexity of the observables.

In this work we focus on the experimental implementation of informationally complete (IC) measurements. IC measurements offer several benefits such as allowing for the estimation of multiple observables from the same measurement data \cite{D_Ariano_2004ICMeasurementsGroups,dariano2007optimal, huang2020predicting, elben2022randomized, dutt2023practical}, a feature that has been proven useful for measurement-intensive algorithms such as ADAPT-VQE \cite{nykanen2023mitigating}, qEOM \cite{morrone2024estimating} and SC-NEVPT2 \cite{fitzpatrick2024quantum}, and providing a seamless interface between quantum and classical hardware, enabling the implementation of efficient error mitigation methods \cite{filippov2022matrix, filippov2023scalable}. But most importantly for the current study, IC measurements allow us to mitigate detector noise by performing quantum detector tomography \cite{Maciejewski2020, Glos2022} and using the noisy measurement effects to build an unbiased estimator for the molecular energy. Furthermore, it also allows us to use efficient post-processing methods that enhance measurement accuracy and decrease shot overhead \cite{malmi2024enhanced, fischer2024dual-frame, caprotti2024optimising, mangini2024low-variance}. Estimation of quantum chemical observables is also possible via non-IC measurements such as Pauli grouping \cite{kandala2017hardware-efficient, izmaylov2020unitary, verteletskyi2020measurement, huggins2021efficient, crawford2021efficient, miller2022hardware}, which allows for the observable to be measured in groups of Pauli strings instead of individual Pauli strings on the quantum computer.

Another technique we use to tackle shot overhead is the implementation of locally biased random measurements \cite{hadfield2022measurements}. This technique allows us to choose measurement settings that have a bigger impact on the energy estimation, reducing the number of shots required, while maintaining the informationally complete nature of the measurement strategy. Circuit overhead, another critical issue, is addressed through repeated settings and parallel Quantum Detector Tomography (QDT). These techniques optimize the use of quantum resources, allowing for more efficient circuit execution. Finally, we observe that the temporal variations of the detector might pose a barrier for achieving high precision measurements on current quantum hardware. To mitigate that, we introduce a blended scheduling technique to account for the dynamic noise. 

Our strategies offer practical solutions to the challenges faced by current quantum computers in making precise and reliable measurements. By integrating these techniques, we can significantly enhance the accuracy of quantum computations, paving the way for more impactful applications of quantum computing technology. In the following sections, we provide detailed explanations and evaluations of each strategy, demonstrating their effectiveness in improving measurement reliability and precision on current quantum platforms.

\section{Results}
\subsection{Case study: Energy estimation of the BODIPY-4 molecule}
For our analysis we consider the Boron-dipyrromethene (BODIPY) molecule. BODIPY and its derivatives are an important class of organic fluorescent dyes that possess attractive chemical and physical features such as high modularity, strong absorption of the visible light, significant fluorescence quantum yields, and exceptional thermal and chemical/photochemical stabilities. Due to these favorable characteristics, the BODIPY compounds has found widespread applications in medical imaging, biolabelling, photoelectrochemistry, photocatalysis, artificial photosynthesis, optoelectronics, and photo-dynamic therapy. 

In this work we study the measurement of an initialization state of a $\Delta$ADAPT-VQE \cite{nykanen2024toward} procedure for an in-solvent BODIPY-4 molecule in various active spaces of 4e4o (8 qubits), 6e6o (12 qubits), 8e8o (16 qubits), 10e10o (20 qubits), 12e12o (24 qubits) and 14e14o (28 qubits).
We provide an estimate of the energy of the ground state ($S_0$), first excited single state ($S_1$) and first excited triplet state ($T_1$) of the BODIPY-4 molecule in each of the selected active spaces. To estimate the excited state energies, we first use the techniques presented in Ref. \cite{nykanen2024toward} to generate Hamiltonians for which the original excited states become ground states, and then use the Hartree-Fock states of these generated Hamiltonians.

\begin{table}[h!]
    \centering
    \begin{tabular}{c|c c c c c c}
        Qubits & 8 & 12 & 16 & 20 & 24 & 28 \\
        \hline
        Num. Pauli strings & 361 & 1819 & 5785 & 14243 & 29693 & 55323  
    \end{tabular}
    \caption{\textbf{The number of Pauli strings contained in the $S_0$, $S_1$ and $T_1$ observables in the various active spaces or system sizes in qubits.} The number of Pauli strings grows as $\mathcal{O}(N^4)$ \cite{cao2019quantum}.}
    \label{tab:num_paulis}
\end{table}
Each Hamiltonian representing the molecule in a given active space has the same number of Pauli strings, which are shown in Table \ref{tab:num_paulis}.

The initialization state is represented by the Hartree-Fock state, which is a separable state and does not require any two-qubit gates to be prepared. This choice comes from the fact that our focus is on the measurement errors alone, so we avoid two-qubit gate errors. Notice however, that the Hamiltonians contain a large number of Pauli strings and have a complex structure, which makes the measurement of such Hamiltonians to chemical precision (\ie $1.6\times 10^{-3}$ Hartree) not trivial even for the Hartree Fock state.

All three Hamiltonians are measured on the initialization state in a blended way, meaning that we have three sets of Hamiltonian-circuit pairs which are executed alongside circuits for QDT in a blended way such that we can assume that each experiment is performed with the same average measurement.
In addition, the goal of the $\Delta$ADAPT-VQE is to estimate gaps between the different energies, which requires the estimations of the energy to be as homogeneous as possible. While we cannot ensure that the overall noise caused in the circuits is the same for all, through blending we can ensure that any temporal fluctuations in the noise is contained evenly in all circuits.
The implementation of blending on IBM computers as well as details regarding executions is described in \suppref{III}.
To assess the quality of the Hamiltonian-inspired locally biased classical shadows, in practice we perform different measurements for the three Hamiltonians, which means that even if the state that is prepared is the same for the three cases, the measurement is different and consequently the experiment performed on the quantum computer is as well.

In the results below, we present two types of errors: Absolute errors and standard errors.
The absolute error represents the distance of the estimated energy $E_{\textrm{est}}$ to some reference energy $E_{\textrm{ref}}$, $|E_{\textrm{est}} - E_{\textrm{ref}}|$.
This error cannot be calculated for cases where the value $E_{\textrm{ref}}$ is not accessible and is merely used as a tool for verification in this work.
The absolute error indicates the accuracy of an estimated energy and can therefore be used to check for the presence of systematic errors.
On the other hand, the standard error is computed as the square root of the estimator variance, which is shown in Eq.~\eqref{eq:repeated_settings_variance} for the repeated settings estimator.
The standard error is a measure of the likely distance between the mean taken from a certain number of samples (shots) and their true ``infinite statistics'' population mean and therefore indicates the precision of the estimation, \ie the presence of random errors.
The measure essentially signifies how confident we can be that our estimate is close to the population mean. 
Different ranges in the confidence are indicated by multiples of the standard error or standard deviation $\sigma$, \eg with a confidence of $68\%$ the population mean is within $1\sigma$, $95\%$ the population mean is within $2\sigma$ and $99.7\%$ the population mean is within $3\sigma$.
In the case when the estimator is plagued by some systematic error or ``bias'' through \eg imperfect measurements, absolute errors should be significantly larger than the standard errors, \ie the absolute error of the estimate cannot be explained by random errors alone.

\subsection{Reduction of estimation bias using QDT}
\label{subsec:bias_reduction_results}

\begin{figure}[h!]
    \centering
    \includegraphics[width=0.95\linewidth]{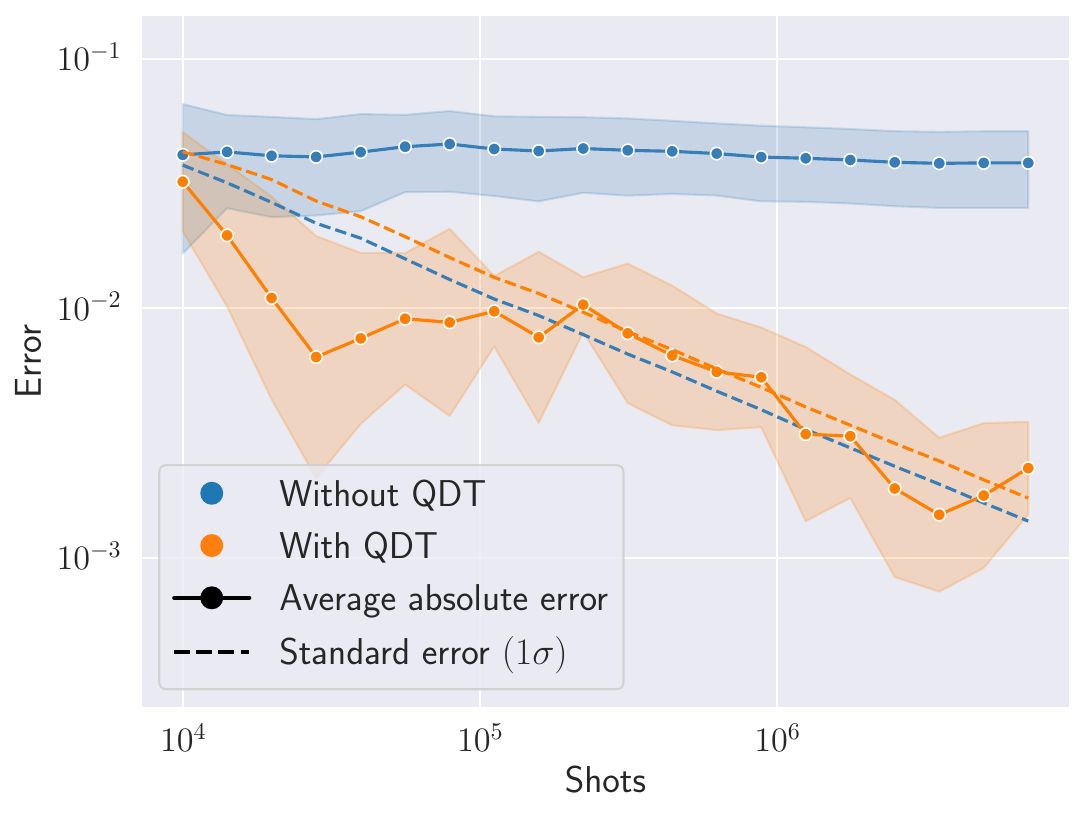}
    \caption{\textbf{Errors obtained for the estimation of the 8-qubit $S_0$ Hamiltonian on the initialization state run on \texttt{ibm\_cleveland}.} The curves indicate energies estimated using POVM effects with and without QDT as a function of the number of shots, averaged over 10 repetitions. The error bars represent the spread of the absolute errors over the 10 repeated experiments with a 95\% confidence interval. The absolute error is computed as the absolute difference to the noiseless value of the initialization energy. The standard error is computed using the expression for the variance for repeated settings in Eq.~\eqref{eq:repeated_settings_variance}.}
    \label{fig:bias_reduction_average}
\end{figure}

To demonstrate the effect of QDT, we measured the 8-qubit $S_0$ Hamiltonian on the initialization state on \texttt{ibm\_cleveland}, performed blended QDT alongside it and repeated the experiment 10 times.
In each repetition we sampled $S=7 \times 10^{4}$ different measurement settings and repeated each setting for $T=100$ shots for a total of $7 \times 10^{6}$.
For the submission of one repetition, we repeated each QDT circuit 4 times per job, of which there were 278 in total.
Each QDT circuit was also measured with 100 shots, adding to a total of $1.11 \times 10^{5}$ shots per QDT circuit and $1.33 \times 10^{6}$ shots for QDT in total.
One repetition took between 45-50 minutes to complete.

We can see from the results that, noise in the measurement process can greatly deteriorate the quality of the energy estimation as shown in Fig.~\ref{fig:bias_reduction_average}.
If we assume that the measurement that was implemented on the quantum computer was described using the ideal effects, the average absolute error with respect to the true initialization energy is $3.83\times 10^{-2}$ and remains mostly unchanged even if more statistics are collected. 
However, if we use QDT to characterize the POVM effects on the device, we reduce the bias to $2.72\times 10^{-3}$ for the same data.
The average standard error over the 10 repetitions for the estimation with blended QDT is $1.75\times 10^{-3}$. 
Whereas it is clear that without QDT, the estimation of the energy incurs biases on the order of $10^{-2}$, when QDT is used to obtain noisy effects, the estimator obeys the $1/\sqrt{S}$ scaling of the standard error and is able to reach errors close to chemical precision ($1.6 \times 10^{-3}$) consistently. 

\subsection{Measurement overhead of the measurement schemes}
\label{subsec:measurement_overhead_results}
\begin{figure}[h!]
    \centering
    \includegraphics[width=0.9\linewidth]{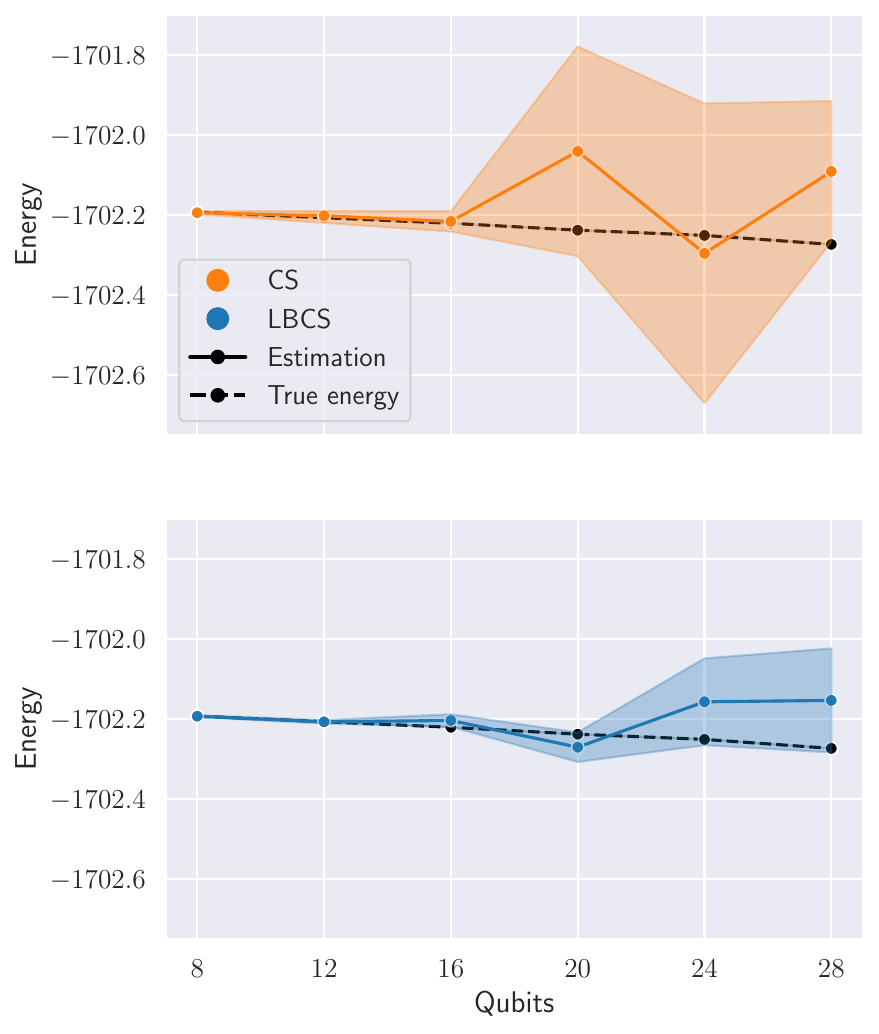}
    \caption{\textbf{Comparison of the CS and LBCS measurement schemes for the measurement of the initialization state for the $S_0$ Hamiltonian in the various system sizes.} The plots include a black dashed line, which indicates the exact initialization energy as a reference. The shaded region represents the standard error, computed using the variance of the repeated settings estimator in Eq.~\eqref{eq:repeated_settings_variance}.}
    \label{fig:povm_energy_scaling_comparison}
\end{figure}
\begin{figure}[h!]
    \centering
    \includegraphics[width=0.9\linewidth]{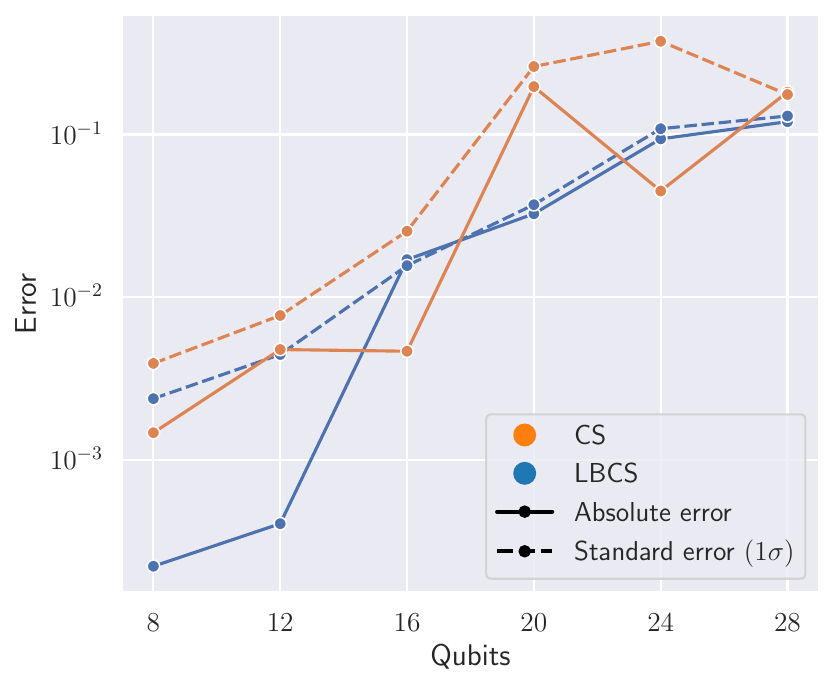}
    \caption{\textbf{Comparison of the estimation errors of the LBCS and CS measurement schemes for the measurement of the initialization state for the $S_0$ Hamiltonian in the various system sizes.} The plot shows the absolute and standard errors ($1\sigma$) of both the LBCS and CS measurement schemes with parallel QDT. The absolute error is computed as the absolute difference to the noiseless value of the initialization energy. The standard error is computed using the expression for the variance for repeated settings in Eq.~\eqref{eq:repeated_settings_variance}.}
    \label{fig:povm_error_scaling_comparison}
\end{figure}
To demonstrate the viability of the measurement techniques presented in this work, we show results of experiments performed for the  Hamiltonians in all the system sizes.
We have in total run twelve separate experiments on \texttt{ibm\_cleveland}, where we measure each of the three Hamiltonians on the initialization state in parallel for all six active spaces that we have considered using both the Classical Shadows (CS) and Locally Biased Classical Shadows (LBCS) measurement strategies.
Each Hamiltonian was measured using $S = 3\times 10^{4}$ measurement settings with $T=100$ shots each.
Alongside the measurement of the three Hamiltonians, we also performed QDT in a blended way, where each of the total 341 jobs contained 3 repetitions of each QDT circuit of the total twelve QDT circuits with each instance being measured for 100 shots, summing to a total of $1.02 \times 10^5$ shots per QDT circuit and $1.23 \times 10^6$ shots for the full QDT procedure.
One run for one set of Hamiltonians for one active space size and measurement strategy took around 1 hour.
A comparison between the LBCS and CS measurement schemes for the measurement of the $S_0$ Hamiltonian initialization energy is shown in Figs.~\ref{fig:povm_energy_scaling_comparison} and \ref{fig:povm_error_scaling_comparison}. The results shown in the figure are obtained with effects characterized using QDT. Additional results for the $S_1$ and $T_1$ Hamiltonians are shown in \suppref{VII}.

We show that QDT reduces errors in the estimation of the energies significantly also for these results in \suppref{VI}.
We can see that both measurement schemes contain the true energy within approximately $1\sigma$ of their error bars, however the LBCS scheme maintains a smaller standard error for all problems in the different system sizes compared to the CS scheme, meaning that on average LBCS will produce estimations with smaller variance.
Despite that, we notice that for both measurement strategies the standard errors, highlighted by the shaded area, increase significantly with the number of qubits.
The large variance presents a major problem for high precision experiments on quantum computers and is caused by both the measurement scheme and the measurement noise levels on the quantum computer.
The effect of the increased variance due to QDT is discussed further in \suppref{VI}.

\subsection{Characterization of temporal noise with blending}

Lastly, we present an example of temporal noise that we have observed in one of the experiments shown in Sec.~\ref{subsec:bias_reduction_results} in more detail.
The experimental setup is identical to the one described in the aforementioned section with the exception that we have also run an additional job with only the twelve QDT circuits and $1\times 10^5$ shots each, a total of $1.2\times 10^6$ shots.
This additional job represents a QDT experiment performed without blending, as it is fully performed before and thus independently from the experiment for the initialization state.
In the presence of temporal noise, we expect the noise to be different between the two time periods in which the non-blended QDT experiment and the initialization state experiment are performed and therefore find a mismatch between the measurement that is implemented in both.
In total, the experiment runtime was around 50 minutes and was performed on \texttt{ibm\_cleveland}.

The temporal noise that we have observed in this particular experiment arose on the qubit with index 1 on the device.
We noticed the temporal noise by marginalizing the frequencies coming from the measurement of the initialization state in the Pauli $Z$ basis throughout the full experiment on this qubit.
This way, we can obtain the frequencies coming from the measurement of the same state with a fixed measurement, which should stay constant throughout the experiment within statistical fluctuations, separately for each of the 278 jobs.

\begin{figure}[h!]
    \centering
    \includegraphics[width=0.9\linewidth]{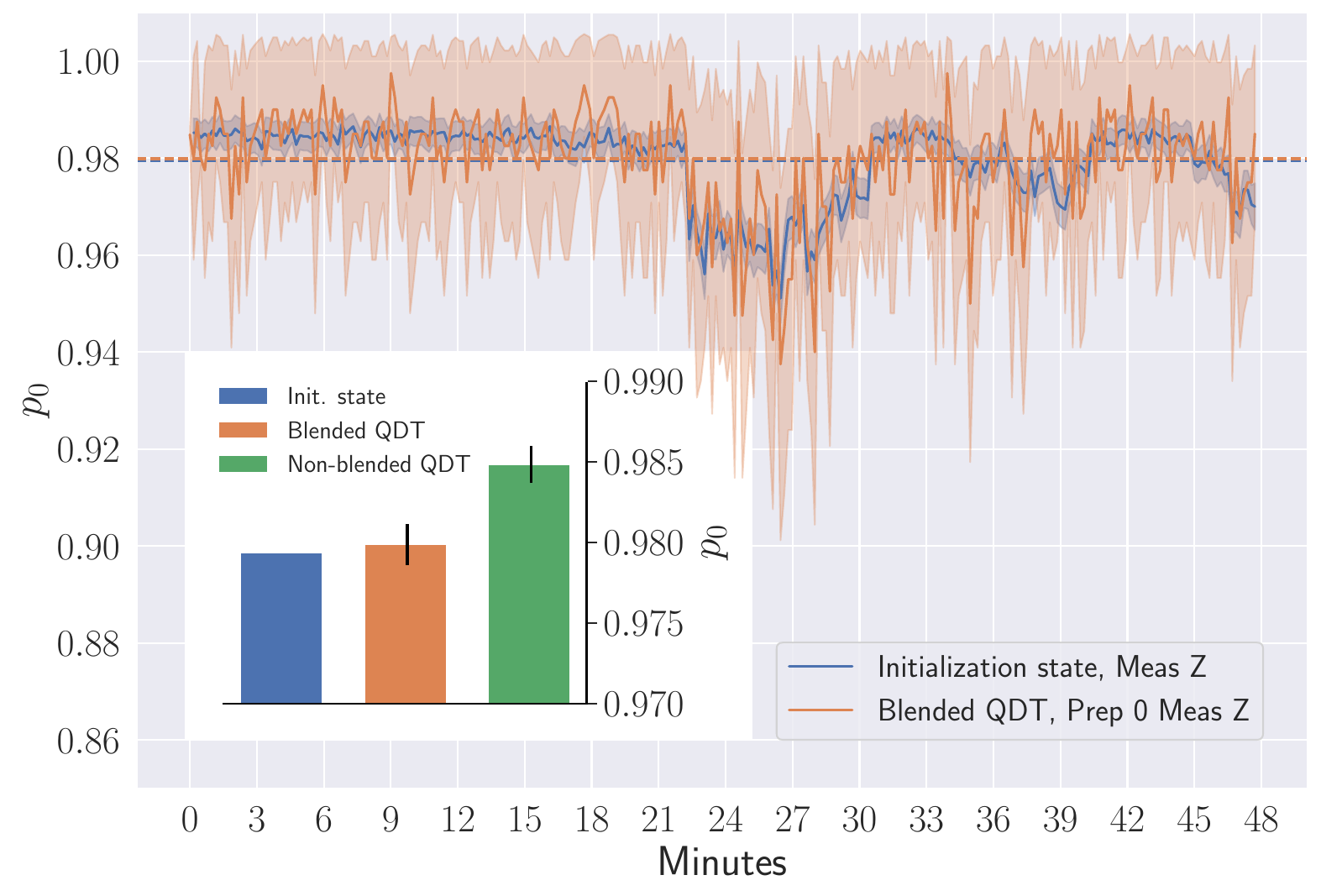}
    \caption{\textbf{Comparison between the frequencies obtained throughout the measurement of the initialization state in the $Z$ basis and the QDT experiment with the measurement of $\ket{0}$ in the $Z$ basis on one qubit.} Each data point in the plot represents the total frequency of obtaining the outcome zero in one job. The line plot shows that the frequencies drop temporarily between the 22nd and 30th minute after which some fluctuations can still be observed. In the inset bar plot, we show the average frequencies from the initialization state as well as blended and non-blended QDT experiments from their complete runtime. All error bars are to $3\sigma$ with the the standard error for binomial random variables, $\sigma = \sqrt{p_0(1-p_0)/S}$.}
    \label{fig:tls_noise-blending}
\end{figure}

In the line plot found in Fig.~\ref{fig:tls_noise-blending} we compare the frequencies obtained by measuring the initialization state in the Pauli $Z$ basis on qubit 1, where $\ket{0}$ is prepared, to the frequencies from blended QDT of the Pauli $Z$ on qubit 1, where $\ket{0}$ is also prepared.
Each data point corresponds to the frequency obtained from all the circuits in one job, which is inherently blended, and is sorted by the execution time of said job.

We can see from the figure that the experiment faces a temporary regime of higher measurement errors, which are statistically significant as indicated by the error bars of the frequencies from the initialization state measurement.
The frequencies arising from blended QDT have larger error bars, due to the fact that we collect fewer samples per job for QDT than the initialization state, however we can see that its curve follows the temporary drop in the measurement frequency.

This is illustrated more clearly by the inset of Fig.~\ref{fig:tls_noise-blending}, which shows the total frequency obtained in both of these datasets.
The frequencies from both experiments match within error bars, which indicates that blended QDT captures the total measurement error, including the temporal noise, well.
In addition, we show the total frequency obtained in the first job, where we collected statistics for the non-blended QDT procedure.
The total frequency arising from the non-blended QDT experiment is much higher than that from the initialization state, suggesting that the effects we would obtain using these results would be different from the ones that were implemented during the initialization experiment, leading to errors in the experiment that are not taken into account.
For this example, the temporal noise causes the frequencies between the initialization state experiment and the non-blended QDT experiment to differ by around $5 \times 10^{-3}$, whereas the blended QDT procedure obtains the frequency with an error that is a magnitude of order smaller, $5 \times 10^{-4}$.

Temporal noise is hard to detect through calibrations as they occur and affect the experiment during its runtime.
We have observed that qubits that exhibit smaller readout errors in their calibration data tend to also have fewer or weaker occurrences of temporal noise.
However, for most applications that are deemed to be interesting on quantum computers, we cannot simply choose qubits purely based on their readout error calibration data, as for more complex circuits one has to take into account two-qubit gate errors as well as connectivity.
In addition, the more qubits are involved in an experiment, the more likely we are to observe temporal noise during it, which is why we believe that blending is a simple but crucial method for high precision measurements.

\section{Discussion}
\label{sec:conclusions}
In this work we have shown how high precision measurements can be performed on current quantum computers in a practical way despite high average readout errors and complex observables. We have discussed numerous limitations of the quantum hardware and presented methods to overcome the challenges caused by them. 

First, we proposed a simple change to the randomized measurement scheme called \textit{repeated settings}, which can be used to reduce the amount of measurement settings that need to be sampled, which helps in reducing pre- and post-processing resources for experiments. To reduce the measurement overhead when estimating expectation values of complex Hamiltonians, such as the ones for the BODIPY-4 molecule presented in this work, we employed a simple and efficient type of locally biased classical shadows that tailors the measurement to the Hamiltonian.

Implementations of these measurements on real quantum computers will incur noise, which introduces errors into the estimation of energies done with them.
For static measurement noise we apply parallel QDT, which on top of reducing measurement errors, allows us to efficiently allocate our pre- and post-processing resources in a similar way to repeated settings.
During our experiments we have also observed temporal measurement noise, which can present a threat to the aforementioned QDT method.
For this purpose, we proposed \textit{blending}, which allows for QDT to take into account both static and temporal noise by executing it in parallel to the experiment used to estimate the expectation value.
All of these mentioned methods contribute to the implementation of measurements with high precision.

We acknowledge the existence of other types of noise such as correlated measurement noise, also termed cross-talk, which parallel QDT does not take into account.
In practice it may be possible for measurement errors to be correlated among qubits \cite{maciejewski2021modeling}, in which case parallel QDT is not sufficient as a method to fully characterize the measurement process.
The effect of cross-talk on estimations of observables using IC-POVMs and ways to mitigate it should be studied and we propose it as a subject of future work.
We do however note that the measurement strategies that we employed in our experiments were enough to achieve high precision, meaning that the effects of cross-talk were not strong enough to cause visible biases in our energy estimations.

In addition, we have applied our methods to system sizes of up to 28 qubits to study the scaling of the noisy measurement process on real quantum hardware. We note that the implementation of the techniques presented in this work will have little to no overhead with the number of qubits and are scalable in nature. 
However, the errors reported for the Hamiltonians beyond 16 qubits are very large for applications in quantum chemistry.
We see two main causes for this increase in variance: The estimator and the measurement errors.
To mitigate the effect of the variance in the estimator, one can explore the use of different measurement schemes, by choosing different POVMs as proposed in Refs.~\cite{Garcia-Perez2021, hadfield2022measurements, hillmich2021decision}. We note that the locally biased random measurements used for the measurement were only a simple heuristic implementation and can most definitely be improved upon. The precision and accuracy of these results may also be improved by post-processing methods such as dual optimization \cite{malmi2024enhanced} and TN-ICE \cite{mangini2024low-variance}. These post-processing techniques are fully compatible with the techniques we use in this paper, given that we use an accurate description of the measurement operators. The measurement errors that we have observed on hardware cause a significant measurement overhead at large qubit numbers, which can make high-precision measurements at a large scale challenging to implement. On one front, the gradual improvement in quantum hardware and their measurement fidelities will alleviate this overhead. On another front, the reduction of measurement errors at such large qubit numbers could be studied by exploring different measurement noise suppression strategies.
For instance, other measurement error mitigation techniques such as T-REX \cite{vandenberg2022model-free}, M3 \cite{nation2021scalable} and ones proposed for sparse probability distributions in Ref.~\cite{yang2022efficient} may be applied to the IC-POVM formalism in conjunction with the techniques we have used in this paper. In addition, techniques based on quantum optimal control \cite{koch2022quantum}, which tailor control pulses based on noise, may improve the quality of the basis rotations performed before the computational basis measurement and Pauli twirling of the native measurement performed in T-REX may reduce the effect of measurement errors and make it easier for quantum detector tomography to obtain accurate and precise measurement operators.

Another natural next step for this research would be to study the performance and robustness of these techniques on more complex states and circuits \eg ones produced by $\Delta$ADAPT-VQE. The study of such complex circuits on hardware will incur a significant amount of noise beyond the measurement process, which will need to be dealt with in some form using noise suppression \cite{koch2022quantum}, noise mitigation techniques \cite{vandenberg2023probabilistic,filippov2023scalable} or error correction \cite{nielsen2010quantum} among others.
The findings in this work will also benefit experimental implementations of algorithms that require informationally complete data such as the previously mentioned ADAPT-VQE \cite{nykanen2023mitigating}, qEOM \cite{morrone2024estimating} and SC-NEVPT2 \cite{fitzpatrick2024quantum}. The techniques may also improve the experimental estimation of properties of quantum states as has been studied using simulations in Ref.~\cite{huang2020predicting}

\section{Methods}

\subsection{Unbiased estimators through informationally complete measurements}
\label{sec:ic_povms}

Any quantum measurement is defined by a positive operator-valued measure (POVM), defined by a set of positive operators $\{\Pi_i\}_{i=1}^r$ called \emph{measurement effects}. Each effect is associated with a possible experimental result (or outcome) $i$ so that the probability of obtaining the $i$th result when measuring the state $\rho$ is given by the Born rule $p_i = \Tr[\rho \Pi_i]$ \cite{nielsen2010quantum}. The effects satisfy $\Pi_i\geq 0$, $\sum_{i=1}^r \Pi_i =\mathbb{I}$, which guarantees that $p_i\geq0$ and $\sum_i p_i=1$.

In this work, we consider a special class of measurements, which are those provided by informationally complete (IC) POVMs, for which the measurement effects form a basis in the space of bounded operators on the system Hilbert space, $\mathcal{B}(\mathcal{H})$. This means that any operator $O$ in this space can be written as $O=\sum_i \omega_i \Pi_i$, for some real coefficients $\omega_i$. As a consequence, the expected value of $O$  can be obtained as 
\be
\label{eq:meanValueDecompPOVM}
\mean{O}=\Tr[\rho O]=\sum_i \omega_i \Tr[\rho\Pi_i]=\sum_i \omega_i p_i,
\ee
for any quantum state $\rho$. This allows us to define the following unbiased estimator of $\mean{O}$:
\be
\overbar{O}=\sum_i f_i \omega_i.
\ee
This realisation provides a powerful way of estimating the expected value of any observable in post-processing: one can simply measure an IC-POVM, obtain the experimental frequencies $\{f_i\}$, and estimate the expected value of any observable as an average of its coefficients $\{\omega_i\}$ over the experimental frequencies \cite{D_Ariano_2004ICMeasurementsGroups}. Moreover, one can also optimize the choice of the IC-POVM in order to minimize the statistical error on $\langle O \rangle$ as a function of the number of experimental shots \cite{Garcia-Perez2021,Nguyen_2022OptShadowTomography,Glos2022}.

In the particular application we consider here, we focus on observables defined on $N$ qubits. In this case, the POVM we consider consists of a product of local IC-POVMs with effects $\{\Pi_{i}^{(q)}\}_{i=1}^r$ on each qubit $q$, where $q=1,\ldots,N$ labels each qubit. We assume that the POVM is either complete or overcomplete (\textit{i.e.},~$r\geq4$ for a single qubit) and each local POVM on each qubit has the same number of effects. In this case any operator $O$ can be decomposed as:
\begin{align}
\label{eq:obsDecompPOVM}
    O &= \sum_{i \in \{1,\dots,r\}^N} \omega_{i} \bigotimes_{q=1}^N \Pi_{i_q}^{(q)},
\end{align}
where $i$ is a multi-index of $N$ local measurement outcomes and $\omega_{i}$ are suitable coefficients of the linear decomposition and efficiently computable classically~\cite{Garcia-Perez2021, mangini2024low-variance}. 

The local POVMs we used in our experiments consist of randomly chosen Pauli measurements in directions $X$, $Y$ or $Z$ on every qubit. In this case, each single qubit measurement has $r=6$ possible outcomes, corresponding to the measurement effects
\ba\label{eq:local-effects}
&\{&p_X\ketbra{+}{+},p_X\ketbra{-}{-},p_Y\ketbra{+_y}{+_y}, p_Y\ketbra{-_y}{-_y}, \nonumber \\
&&p_Z\ketbra{0}{0},p_Z\ketbra{1}{1}\},
\ea
where each projector corresponds to a projection on an eigenstate of the Pauli operators, and $p_X$, $p_Y$, and $p_Z$ correspond to the probabilities of picking each measurement direction. The measurement schemes we use in this work are particular instances of dilation-free IC-POVMs presented in Ref.~\cite{Glos2022}. In \suppref{I} we discuss how to improve the estimation precision by choosing the probability distribution of the local random measurements based on an approach introduced in Ref.~\cite{hadfield2022measurements}. 

\subsection{Repeated measurement settings for reducing circuit overhead}
\label{subsec:repeated_settings}
In practice, dilation-free POVMs are implemented as a series of randomly sampled measurement bases, represented by different basis rotation circuits that are appended to the end of the state that is being measured.
Thus, a single shot involves both the sampling of a measurement basis as well as the measurement of the state in question in this measurement basis.
Since the total number of possible measurement settings is $r^N$ for a $N$-qubit POVM with $r$ effects per qubit, the number of different circuits to run on the quantum device will be essentially equal to the number of shots for large system sizes.

Many quantum computers, such as IBM's quantum computers, set a limit to how many circuits may be submitted in one job, partly due to limitations in the electronics of the device.
Submitting a large number of unique circuits can incur additional classical pre- and post-processing overhead and is ideally avoided as much as possible. 

We propose a simple change to the dilation-free POVM measurement scheme to reduce the number of unique circuits that need to be submitted, while providing unbiased estimates of the mean and variance.
For dilation-free POVMs we may consider the number of measurement settings $S$ and the number of shots allocated for each measurement setting $T$ separately. We label the outcome $j$ obtained through the measurement setting $i$ as $m_{i,j}$. The Monte Carlo estimator $\overbar{O}$ for dilation-free POVMs is then
\begin{equation}
\label{eq:repeated_settings_estimator}
    \overbar{O} = \sum_{i=1}^S\sum_{j=1}^{T} \frac{\omega_{m_{i,j}}}{ST}.
\end{equation}

The estimator in Eq.~\eqref{eq:repeated_settings_estimator} represents the average of $\omega_m$ over all the sampled measurement settings and their outcomes.
Notice how for $T = 1$, we recover the estimator for the standard one-shot-per-setting dilation-free POVM measurement scheme, whereas for $T > 1$ we perform measurements in the same measurement setting multiple times.
We can thus allocate our total shot budget between the number of settings $S$ and shots per setting $T$ and avoid additional overhead caused by large values of $S$.
We call this procedure \textit{repeated settings}.

In \suppref{V} we show that the estimator in Eq.~\eqref{eq:repeated_settings_estimator} is unbiased and that the variance can be estimated as in Eq.~\eqref{eq:repeated_settings_variance}:
\begin{equation}
\label{eq:repeated_settings_variance}
    \textrm{Var}[\overbar{O}] = \frac{\langle\langle \omega \rangle_i^2\rangle - \langle \omega \rangle^2}{S} + \frac{\langle \omega^2 \rangle - \langle\langle \omega \rangle_i^2\rangle }{ST},
\end{equation}
where $\langle \omega^{n} \rangle$ denotes the $n$-th moment of the random variable $\omega$ taken over all possible values of $\omega_i$ as defined in Eq.~\eqref{eq:meanValueDecompPOVM}. In practice, the quantity can be estimated from all obtained samples $\omega_{m_{i,j}}$ as defined in Eq.~\eqref{eq:repeated_settings_estimator} by averaging over them. The quantity $\langle\langle \omega \rangle_i^2\rangle$ denotes the \textit{conditional second moment}, which is the second moment of the means obtained for each measurement setting $i$.
In practice, this quantity can be estimated by keeping the results coming from different measurement settings separate, computing the mean squared for each and then taking its mean over all settings.

The conditional second moment is bounded as $\langle \omega \rangle^2\leq \langle\langle \omega \rangle_i^2\rangle \leq \langle \omega^2 \rangle$. For the case where the conditional second moment is at its minimum, \ie $\langle\langle \omega \rangle_i^2\rangle = \langle \omega \rangle^2$, the variance can be reduced purely by increasing the number of shots allocated per measurement setting. On the other hand, when the conditional second moment is at its largest, \ie\ $\langle\langle \omega \rangle_i^2\rangle = \langle \omega^2\rangle$, the variance is affected only by the number of measurement settings, which means that higher precision cannot by attained by repeating settings.
For certain estimators, it is also possible to obtain the same variance for different choices of $S$ and $T$ and thus total number of shots $ST$.

A caveat with this method is that there is a fundamental limit in the precision that can be reached by only increasing the value of $T$, as the repeated settings scheme performs at best as well as the one-shot-per-setting scheme in terms of precision.
We may investigate this limit using the inequality in Eq.~\eqref{eq:saving_factor}. The derivation of the inequality is presented in \suppref{V}.
\begin{equation}
\label{eq:saving_factor}
    S \geq \left(1-\frac{\langle \omega^2 \rangle - \langle \langle\omega\rangle_i^2 \rangle}{\langle \omega^2 \rangle - \langle \omega \rangle^2}\right) R = F_{\textrm{saving}} R,
\end{equation}
where $S$ denotes the number of measurement settings in the repeated settings scheme and $R$ denotes the number of total shots in the one-shot-per-setting scheme.
The quantity $F_{\textrm{saving}} R$ represents the number of measurement settings $S$ needed to achieve the same precision one would obtain with $R$ shots in the one-shot-per-setting scheme in the limit $T \to \infty$.

In practice, it is possible for the repeated settings estimator to reach the same precision with a finite and often low value of $T$ depending on the value of $\langle \langle\omega\rangle_i^2 \rangle$.
The value $F_{\textrm{saving}}$, which we call the \textit{saving factor}, is state, observable and POVM dependent and is bounded by $0 \leq F_{\textrm{saving}} \leq 1$. Ideally, the value $F_{\textrm{saving}}$ would be small such that we may obtain the same precision as in the one-shot-per-setting scheme with only few measurement settings $S$ but with large numbers of shots per setting $T$, as it is often favorable to request many shots per measurement setting than measurement settings themselves.

\subsection{Dealing with noisy detectors}
\label{sec:qdt}
In real applications, measurements are never perfect due to the presence of noise in the device. In this case, the POVM effects are not given by Eq.~\eqref{eq:local-effects}, but by a noisy version of it. In the case of standard projective measurements, several strategies to mitigate measurement error have been proposed \cite{nation2021scalable,vandenberg2022model-free}. In the case of IC measurements, we can use the fact that if the noise level is not too high, the noisy effects are likely to form a IC measurement as well. In this case, we can still use the noisy measurement to obtain unbiased estimators of any observable.  

In our experiments we have used a technique called Quantum Detector Tomography (QDT) to obtain an accurate description of the noisy POVM effects. In particular, we have used the QDT method proposed in Ref.~\cite{Cattaneo2023}, which consists of measuring an informationally complete set of inputs (\ie a set of states that span the Hilbert space such as the eigenstates of the Pauli matrices) and recovering the POVM effects through a likelihood method based on semi-definite programming. One of the main advantages of this method is that it can also incorporate errors in the set of input states. 

A convenient advantage of local single-qubit POVMs is that we may perform QDT on each single-qubit measurement in parallel. This avoids having to run a possibly exponential number of circuits needed to prepare the set of input states required for the QDT procedure, because we can prepare the same input state for each qubit in parallel.
In addition, for dilation-free POVMs the exponential number of measurement settings can be avoided by explicitly performing tomography locally on the set of measurement bases (which are in the case of classical shadows simply the Pauli operators $\{X, Y, Z\}$) without performing the classical sampling procedure of the global measurement bases.
For a fixed set of input states, the number of circuits required for QDT is constant, which is simply the product of the number of input states and the measurement bases.
For instance, if we perform QDT using the input states $\{\ket{0},\ket{1},\ket{+},\ket{+_y}\}$ on an $N$-qubit POVM with three measurement bases each, we only need twelve circuits in total regardless of the value of $N$.
Additionally, the effects of Pauli measurements characterized using a single QDT procedure can be reused for any local classical shadows-style POVM including locally biased classical shadows or any measurement scheme involving measurements in the Pauli basis.

\subsection{Dealing with temporal noise}
\label{sec:blending}
\begin{figure*}[ht!]
    \centering
    \includegraphics[width=0.7\linewidth]{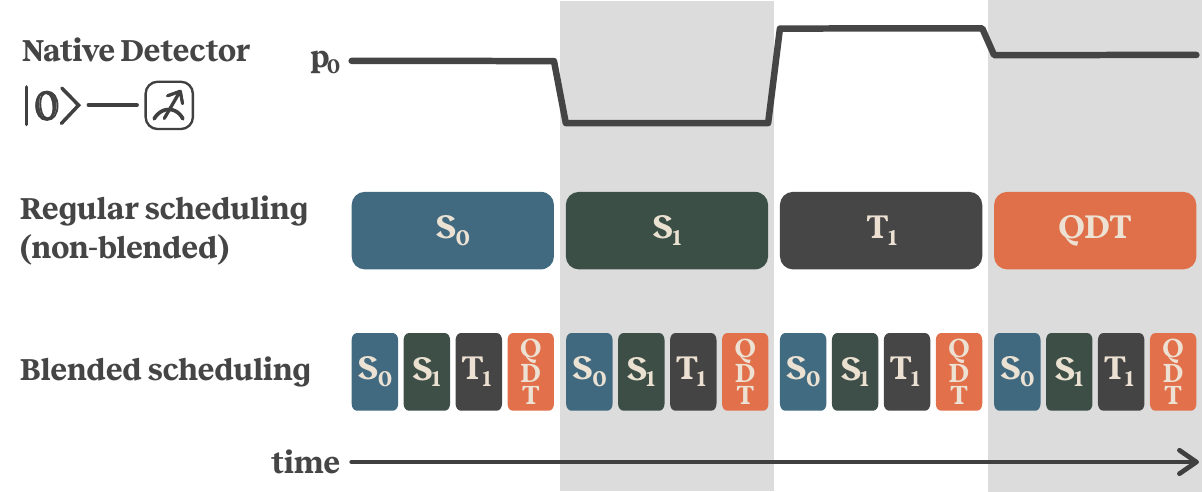}
    \caption{\textbf{A diagram depicting a fluctuating detector as a function of time and two ways of scheduling experiments during the same time frame.} Let us assume that the native detector exhibits temporal fluctuations, which can be identified by the frequency $p_0$ of obtaining outcome 0 when measuring $\ket{0}$ over time. For this example, the detector has four distinguishable regions of the aforementioned frequencies. For our experiment, we want to measure three states $S_0$, $S_1$ and $T_1$ as well as run tomographical circuits for QDT. Naively, we could run these circuits as fully separate batches as depicted by regular scheduling. In the presence of temporal noise, we see that the detector that was used for the measurement of the three states can be different from the detector that was used to obtain results for QDT, and thus we may incur bias in the estimation of the energies of these states due to the mismatch in detectors. However, if we schedule the circuits in a blended way and all circuits are executed within a region where the detector remains constant, we can ensure that the measurement of all states as well as the tomographical circuits for QDT are performed with the same detector, allowing us to avoid potential estimation biases due to a mismatch in detectors.}
    \label{fig:blending-diagram}
\end{figure*}
When one measures a single qubit in the computational basis on a near-term quantum computer, the outcome will be affected by a measurement error that is typically around 1-3\% on the IBM quantum computers. This kind of systematic error can be corrected by computing the so-called assignment or transition matrix by preparing a qubit in the $\ket{0}$ and $\ket{1}$ states and determining their error rates and then applying its inverse to the experimental outcomes or through more refined methods that make use of QDT \cite{Maciejewski2020,maciejewski2021modeling}.

However, a well-known issue of current superconducting quantum computers is the presence of (usually unpredictable) temporal fluctuations that affect, for instance, the single-qubit error rate on local measurements \cite{Kim2023,Hirasaki2023}. This means that the measurement effects of the detector that we are employing to measure in the computational basis is effectively fluctuating in time. In other words, the assignment matrix is not constant, and the effects of the POVM we are implementing change in time. More specifically, the noise for a fluctuating detector may be modeled as a series of ``good'' and ``bad'' temporal regimes characterized by the small random perturbations around respectively a reasonable value of the measurement error (\eg 1-3\% as we said above) and a more severe measurement error (\eg 5-10\%). 

The fact that the detector is fluctuating in time is a serious threat to the measurement strategy based on IC-POVM and QDT we have introduced in Sec.~\ref{sec:ic_povms} and \ref{sec:qdt}. Indeed, if we run the circuits for QDT on real hardware at a different time with respect to the circuits for the estimation of the observable mean value as given by Eq.~\eqref{eq:meanValueDecompPOVM} and the detector on the quantum computer varies during this period of time, then the effects we are reconstructing through QDT do not correspond anymore to the ones we should use in Eq.~\eqref{eq:meanValueDecompPOVM}, and the mean value of the observable may be significantly biased.

A common method of reducing the impact of temporal fluctuations on near-term quantum computers is based on ``filtering'', that is, one discards the outcomes obtained during the ``bad'' regimes using different statistical methods \cite{Kim2023,Hirasaki2023}, so that the effects of a fluctuating detector on the experimental results are restricted to the small perturbations around the ``good value''. Experimental observations show that different qubits on a quantum computer switch between the good and the bad regime during different periods of time. Therefore, for a larger number of qubits we will have less and less time windows in which all the qubits are in the good regime. Thus, filtering out all the bad regimes is not viable for practical purposes. 

We propose a new method to cope with temporal fluctuations that relies on a time average that gives rise to an \textit{unbiased} estimator of the mean value of any observable. We call this procedure \textit{blending}. With blending, one ensures that samples for all experiments are taken evenly with respect to chunks of time on the macroscopic level, \ie throughout the entire experiment.
This allows us to collect data for QDT on the same measurement process that is implemented for any other experiment run alongside it, even if the measurement process exhibits temporal fluctuations.
A diagram depicting an abstract example of a fluctuating detector as well as a comparison between ``regular'' and blended scheduling is shown in Fig.~\ref{fig:blending-diagram}.
The method will produce an unbiased estimator even in the presence of temporal noise at the cost of a higher variance, which depends on the noise levels on the quantum computer.
A mathematical demonstration of blending is shown in \suppref{II}, details about its implementation on IBM computers are provided in \suppref{III}.
We also present a noise model for a fluctuating detector using unbalanced random telegraph noise in \suppref{IV}. Unbalanced random telegraph noise is a sign of non-Markovianity on superconducting qubit systems and has been a topic of research in the theoretical and experimental studies in Refs.~\cite{galperin2006non-gaussian, wold2012decoherence, li2013motional, franco2013spin-echo, daniotti2018qubit}.

Blending, on top of allowing unbiased estimations in the presence of temporal fluctuations, also has the advantage of being a noise monitoring procedure. Because shots from different experiments are obtained evenly throughout the runtime of the full set of experiments, one can track how outcomes, which we expect to remain the same throughout the experiment, fluctuate in time.
In principle, besides monitoring only measurement errors, one could also continuously monitor how outcomes fluctuate during the execution of the circuits themselves.
CNOT gates are also known to induce temporal fluctuations \cite{Kim2023} and by comparing the temporal fluctuations seen in the experimental circuits and the QDT circuits, one can deduce whether the fluctuations are caused by the measurement or the state preparation. This sort of analysis is out of scope for the experiments performed in this work and we suggest it as a subject of future work.

\section*{Data Availability}
The data that support the findings of this article are available on Zenodo \cite{zenododataset}.

\section*{Code Availability}
The code that was used to obtain the data that support the findings in this article cannot be made publicly available because it contains commercially sensitive information.

\begin{acknowledgments}
We thank Laurin Fischer, Francesco Tacchino and Ivano Tavernelli for helpful discussions about experiments on IBM quantum computers and temporal noise.
Work on “Quantum Computing for Photon-Drug Interactions in Cancer Prevention and Treatment” is supported by Wellcome Leap as part of the Q4Bio Program.
Access to IBM Cleveland Clinic System One was provided via Cleveland Clinic Quantum Innovation Catalyzer Program.
We also thank the NKFIH OTKA grant FK 135220.
\end{acknowledgments}

\section*{Author Contributions}
All authors contributed to the conception, design and research of the methods presented in this article. K.K., H.V., A.G., M.C. and M.A.C.R. contributed to the software used to perform the experiments. K.K. and H.V. performed the experiments. K.K., H.V., A.G., M.C., M.A.C.R. and D.C. contributed to the drafting of the manuscript. D.C. supervised the project.

\section*{Competing Interests}
The authors declare no competing interests.

\end{document}